\documentclass[10pt,dvipsnames]{article}
\usepackage[a4paper, margin=1in]{geometry}
\usepackage[utf8]{inputenc}
\usepackage[T1]{fontenc}
\usepackage{graphicx}
\usepackage{amsmath}
\usepackage{float}
\usepackage{authblk}
\usepackage{subcaption}
\usepackage{braket}
\usepackage{cite}
\usepackage{booktabs}
\usepackage{cuted}
\usepackage{xcolor}
\usepackage{hyperref}

\definecolor{darkgreen}{rgb}{0,0.5,0}

\title{Benchmarking Single-Photon Sources for Measurement-Device-Independent Quantum Key Distribution}

\author[1]{Janka Memmen\thanks{Corresponding author: \texttt{janka.memmen@tu-berlin.de}}}
\author[1]{Yvonne Lasars}
\author[2]{Tobias Heindel}
\author[3]{Nathan Walk}
\author[1]{Anna Pappa}

\affil[1]{Electrical Engineering and Computer Science Department,
Technische Universit\"at Berlin, 10587 Berlin, Germany}
\affil[2]{Department for Quantum Technology, Universit\"at M\"unster, 48149 M\"unster, Germany}
\affil[3]{Dahlem Center for Complex Quantum Systems, Freie Universit\"at Berlin, 14195 Berlin, Germany}
\date{\today}

\begin{document}

\maketitle

\begin{abstract}
\noindent Measurement-Device-Independent Quantum Key Distribution (MDI-QKD) enables secure communication without trusting the measurement devices, but its performance critically depends on the quality of the employed photon sources. We compare MDI-QKD implementations based on Weak Coherent Pulses (WCPs) with those employing Quantum Dot (QD) single-photon sources. WCP-based schemes are fundamentally limited by Poissonian photon-number statistics, imposing a trade-off between single- and multi-photon emissions, whereas QD sources exhibit sub-Poissonian statistics with low antibunching parameter $g^{(2)}(0)$. Using a decoy-state analysis for both source types, we find that even at moderate mean photon numbers, QD-based implementations outperform the best-case WCP configuration - where the mean photon number is optimised at each transmission distance - across the full range of transmission distances. We further incorporate the effect of imperfect photon indistinguishability via Hong-Ou-Mandel interference visibilities into the security framework and demonstrate that both source types are highly sensitive to even small deviations from ideal indistinguishability, significantly affecting achievable key rates and maximum transmission distance.

\end{abstract}

\section{Introduction}
Quantum key distribution (QKD) offers information-theoretic security based solely on the laws of physics \cite{Bennett:1984}. 
Over the past decade, QKD has progressed from laboratory demonstrations to real-world deployments, with fiber-based implementations achieving secure key exchange over hundreds of kilometers~\cite{Liu:2023, Boaron:2018} and satellite-based links enabling intercontinental key distribution~\cite{Liao:2018}. However, practical implementations inevitably deviate from idealized theoretical assumptions, introducing vulnerabilities that undermine this promise. The use of non-ideal photon sources exposes implementations to photon-number-splitting (PNS) attacks, while the assumption of trusted measurement devices leaves protocols susceptible to detector side-channel attacks \cite{Fung:2007, Zhao:2008, Lydersen:2010, Gerhardt:2011}. The decoy-state method \cite{Lo:2005_decoy} provides an effective countermeasure against PNS attacks, whereas vulnerabilities associated with untrusted detectors motivated the development of measurement-device-independent QKD (MDI-QKD) \cite{Lo:2012,Pirandola:2012}.

In MDI-QKD, the two communication parties (Alice and Bob) transmit quantum states to an untrusted measurement unit, which performs a Bell-state measurement (BSM). Since the security proof makes no assumptions about the measurement apparatus, all detector side-channel attacks are eliminated by design.
To date, most theoretical analyses and experimental implementations of MDI-QKD rely on weak coherent pulse (WCP) sources, which follow Poissonian photon-number statistics. Such sources are inherently limited by a trade-off between maximizing single-photon emission and suppressing multi-photon contributions, which fundamentally caps both the achievable key rate and the maximum secure transmission distance.

Quantum dot (QD) single-photon sources offer a compelling alternative \cite{Vajner:2022,Heindel:2023}. 
Unlike WCPs, QD sources emit photons with near-deterministic single-photon statistics and strongly suppressed multi-photon contributions, characterized by a low second-order coherence $g^{(2)}(0)$. State-of-the-art QD sources have demonstrated high multi-photon suppression ($g^{(2)}(0)=(7.5\pm 1.6)\times 10^{-5}$~\cite{Schweickert:2018}), as well as high photon indistinguishability (exceeding $98\%$~\cite{ding2025high}), making them increasingly viable for practical deployment. Significant progress has also been made with devices operating directly at telecom C-band wavelengths~\cite{Holewa:2025}, where recent reports demonstrate photon-indistinguishabilities exceeding 90\%~\cite{Hauser:2026,Behrends:2026a}, even at record-high optical clock rates of 2.5\,GHz~\cite{Behrends:2026b}. Experimental demonstrations of BB84 single-photon QKD using QD sources include experiments in deployed optical fibers at the metro-~\cite{Zahidy:2024} and inter-city-~\cite{Wang:2026} scale. QD sources have also been shown to outperform WCP-based implementations in several cryptographic protocols, including experimental QKD~\cite{Zhang:2025} and quantum coin flipping~\cite{Vajner:2026} as well as theoretical studies on twin-field QKD~\cite{Bozzio:2022}, motivating their application to MDI-QKD. Despite recent progress in the experimental implementation of single-photon QKD with a security analysis accounting for device imperfections~\cite{Corrigan:2026}, side-channel attacks remain a serious threat, further motivating the implementation of device-independent concepts.

Beyond photon-source statistics, the performance of MDI-QKD depends critically on the quality of the BSM at the MU, which requires high Hong-Ou-Mandel (HOM) interference visibility between the photons arriving from Alice and Bob. This visibility is determined by the indistinguishability of the interfering photons. However, imperfect indistinguishability can introduce vulnerabilities beyond a reduced BSM success probability, by rendering the quantum states prepared in the $X$ and $Z$ bases partially distinguishable, and therefore opening a side-channel that an eavesdropper can exploit. These type of source flaws are captured by the so called basis imbalance and have been studied in both conventional and MDI-QKD. Previous work has considered both imperfections in the qubit encoding itself under the assumption of no extra distinguishing degrees of freedom \cite{Xu_source_exp:2015,Xu_MDI_source:2015,Tang_MDI_source_exp:2016}, and distinguishability arising from additional photonic degrees of freedom. Ref.~\cite{Pereira:2019} considers both types of source imperfections, but assumes that the extent of the imperfections can be obtained through prior source characterization of specific, potentially distinguishing degrees of freedom. 

Here, we take a different approach and adopt the framework of \cite{Duplinskiy:2021}, in which the HOM visibility is used as an experimentally accessible measure of distinguishability that is sensitive to all photonic degrees of freedom. 
We analyze the performance of MDI-QKD with QD single-photon sources and compare it directly to an equivalent WCP-based implementation. For both source types, we employ the decoy-state method to account for non-zero multi-photon components of realistic sources. While decoy-state methods are well established for WCP-based QKD, their application to sub-Poissonian QD sources has received comparatively little attention in the literature. Recently, a security analysis appeared showing that decoy methods with QD sources can improve performance for standard QKD in relevant parameter regimes, alongside a proof-of-principle experimental demonstration \cite{Barnes:2026}. Although neither the decoy-state nor basis choice modulation were carried out in real-time, this work provides evidence for the feasibility of decoy state protocols with QD sources in the future and motivates their study in other cryptographic settings. We further incorporate imperfect HOM visibility into the security analysis, allowing us to assess the impact of photon indistinguishability on both source types. Our results demonstrate a substantial performance advantage for QD-based implementations, achieving secure key generation over significantly longer transmission distances than their WCP counterparts.

\section{Results}
\label{sim_results}
\begin{figure}[htb]
    \centering
    \includegraphics[width=0.5\columnwidth]{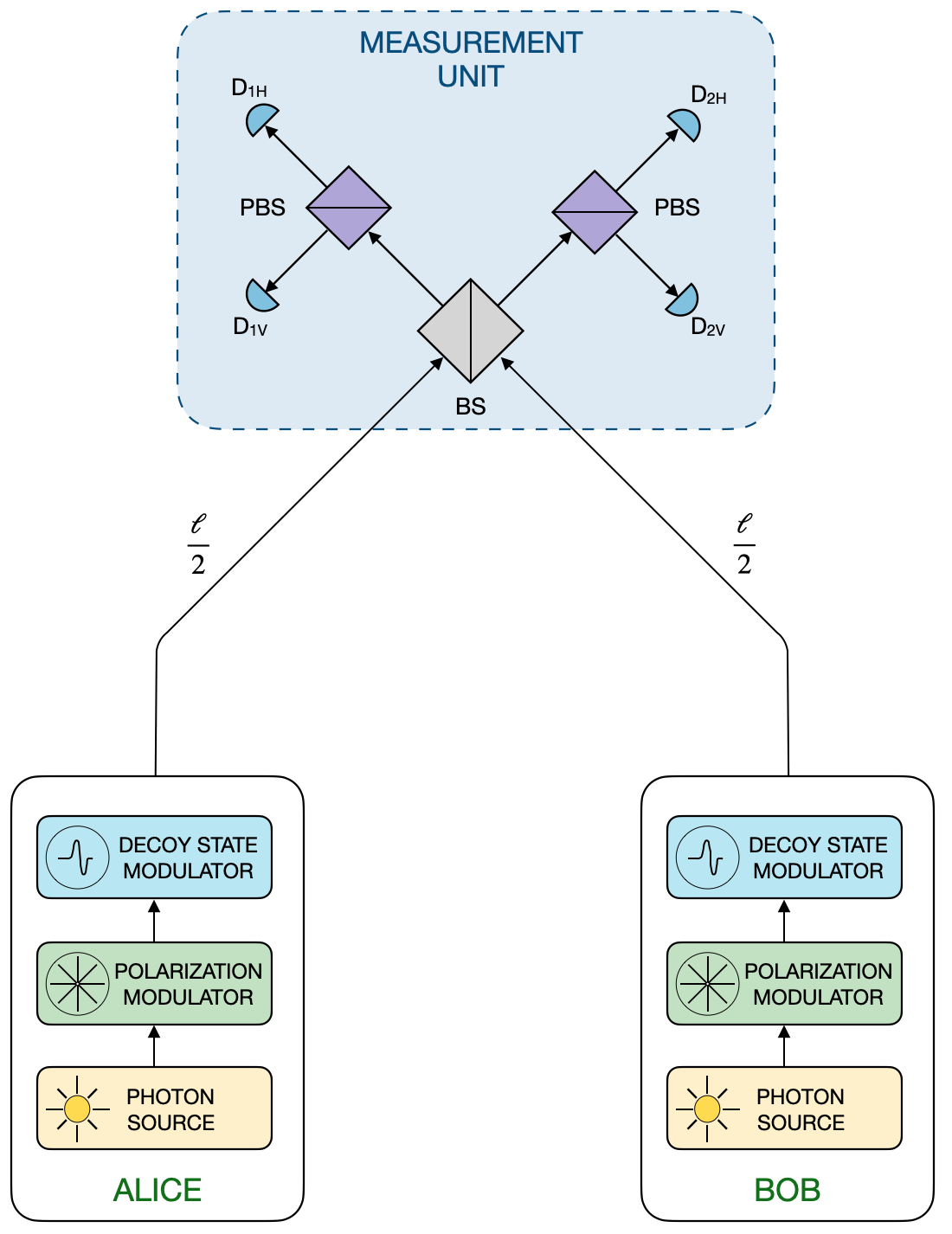}
    \caption{Schematic setup of an MDI-QKD implementation. Alice and Bob independently prepare photons in randomly chosen BB84 polarization states and transmit them to an untrusted measurement unit. There, a Bell state measurement (BSM) is performed via 50:50 beam splitter (BS), polarizing beam splitters (PBSs) and four single-photon detectors. Alice's and Bob's laboratories are trusted, whereas the measurement unit may be fully untrusted. The distance between the two parties and the measurement unit is $\frac{\ell}{2}$, making the total communication distance equal to $\ell$.}
    \label{fig:Setup}
\end{figure}

We consider an MDI-QKD implementation in which Alice and Bob each transmit photons to a central, untrusted measurement unit located midway between them, as illustrated in Fig.~\ref{fig:Setup}. We compare two implementations that differ only in their photon source: one based on WCPs and one based on QD single-photon sources. For both source types, we employ the decoy-state method to bound the contributions of different photon-number components in the security analysis. All other experimental parameters are held identical, including detector efficiency, misalignment error, dark count probability, and fibre attenuation, allowing a direct assessment of the performance advantage conferred by the source statistics alone. For the QD source, we additionally fix the antibunching parameter $g^{(2)}(0)=0.02$ \cite{ding2025high}, which characterizes the degree of single-photon purity.

\begin{figure}[htb]
    \centering
    \begin{subfigure}[t]{0.49\columnwidth}
        \centering
        \includegraphics[width=\linewidth]{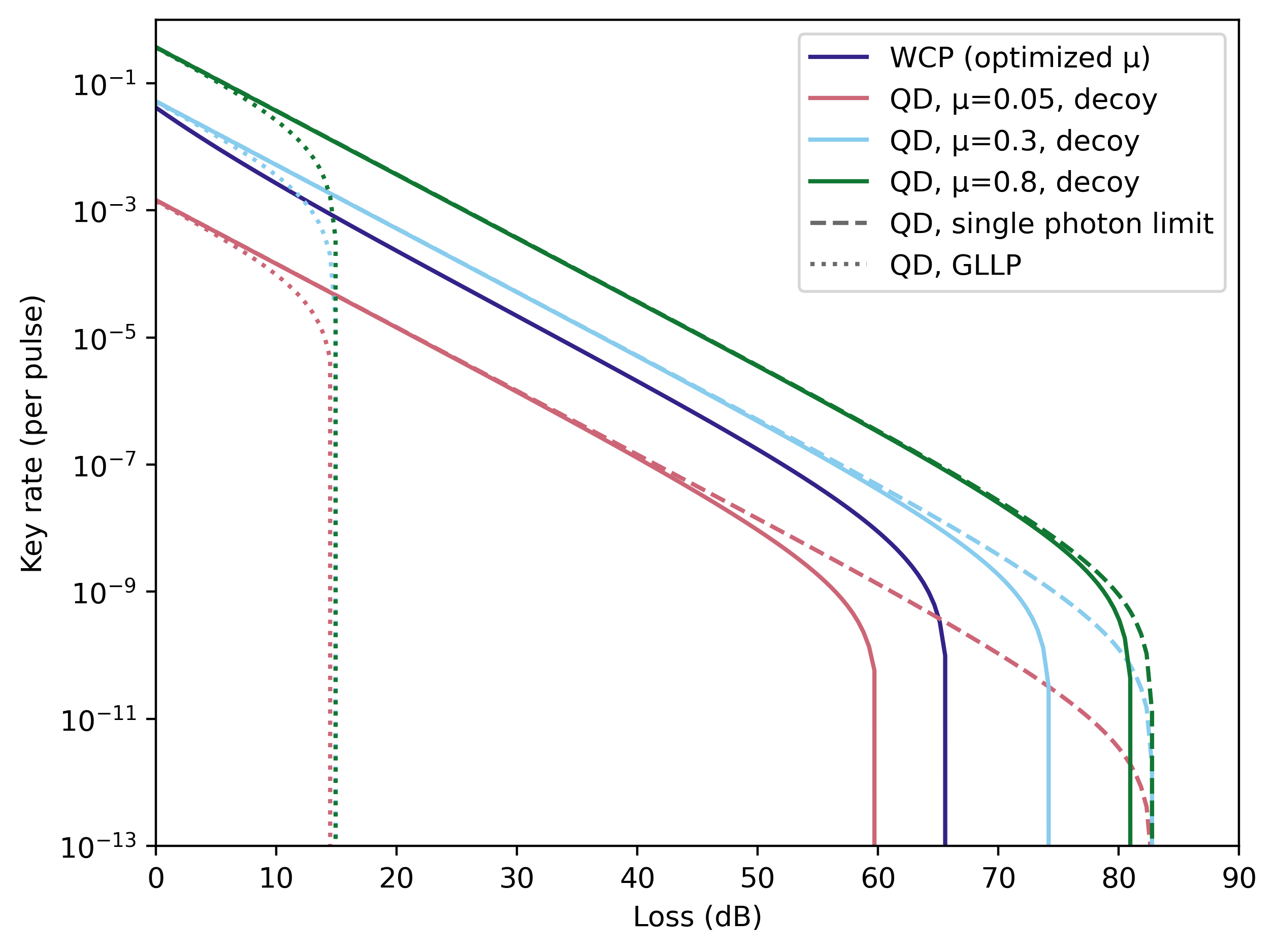}
        \caption{}
        \label{fig:MDI_WCP_QD}
    \end{subfigure}
    \hfill
    \begin{subfigure}[t]{0.49\columnwidth}
        \centering
        \includegraphics[width=\linewidth]{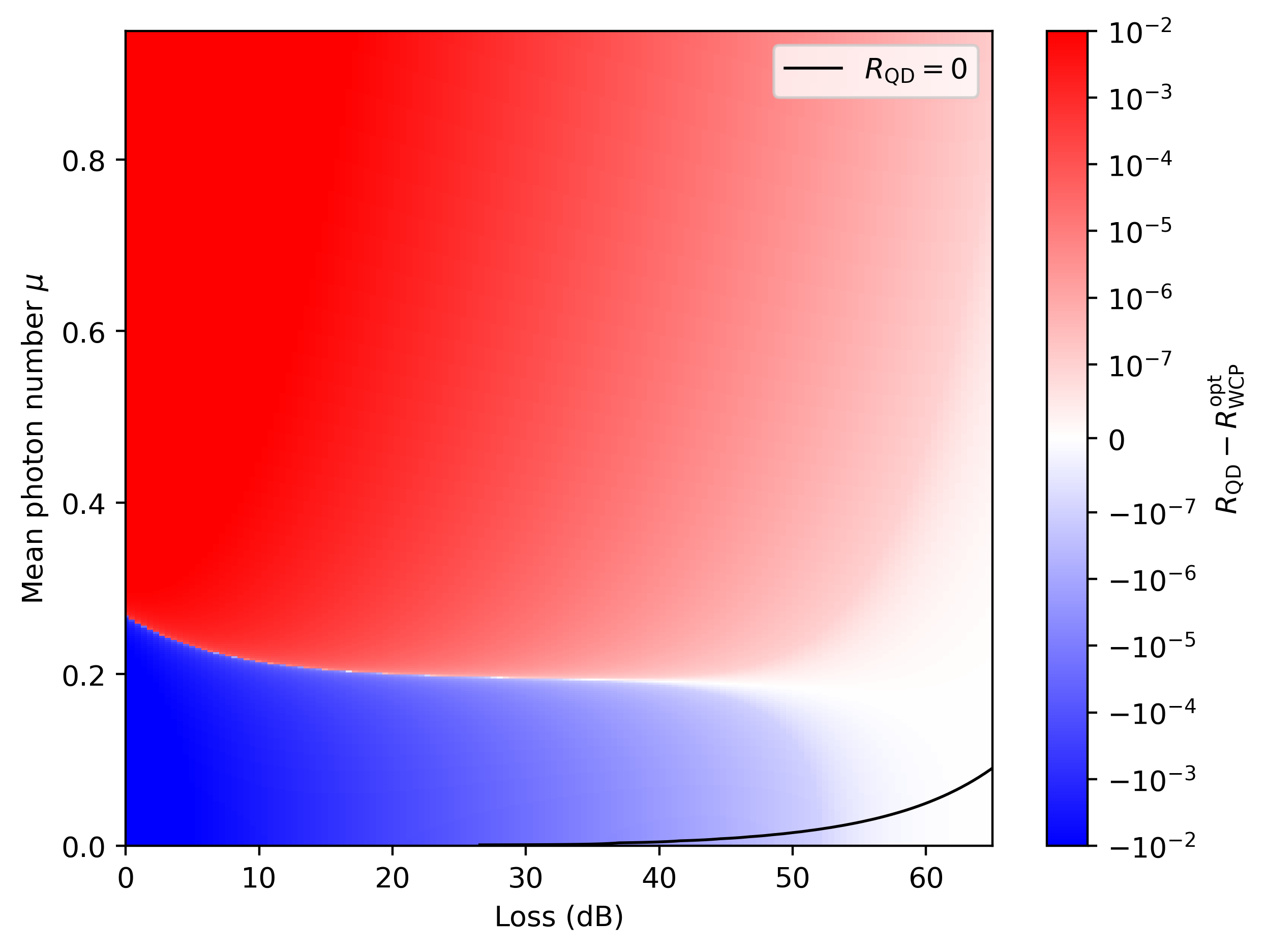}
        \caption{}
        \label{fig:MDI_WCP_QD_2D}
    \end{subfigure}
    \caption{
    Analysis of the asymptotic secret key rates of WCP- and QD-based MDI-QKD implementations, assuming equal loss in the two links between Alice and Bob. In both panels, the mean photon number $\mu$ of the WCP sources is numerically optimized for each total loss, and the mean photon numbers of Alice's and Bob's sources are always assumed to be equal. (a) Secret key rates for WCP sources (solid line) and QD sources using decoy states, as well as QD sources without decoy states, assuming either perfect QD single-photon sources without multi-photon emissions (dashed lines) or the GLLP method\cite{GLLP:2004} (dotted lines), as a function of the total transmission loss. The different colours correspond to different mean photon numbers $\mu$ of the QD sources. (b) Difference $R_{\mathrm{QD}}-R^{\mathrm{opt}}_{\mathrm{WCP}}$ between the key rates of the QD and optimized WCP implementations, both using decoy states, as a function of the total loss and the QD mean photon number $\mu$. The black contour marks the boundary beyond which the QD implementation cannot achieve a positive key rate. Blue regions indicate a performance advantage for WCP sources, while red regions indicate an advantage for QD sources.
    }
    \label{fig:MDI_WCP_QD_comparison}
\end{figure}

Fig.~\ref{fig:MDI_WCP_QD} shows the key rates as a function of the total loss, assuming that the measurement unit is placed midway between Alice and Bob and that the two links have symmetric losses. The corresponding methods are described in Section~\ref{key_rate}.

For the WCP implementation, the mean photon number $\mu$ is numerically optimized at each transmission loss. In contrast, for the QD source we consider the fixed values $\mu\in\{0.05,\,0.3,\,0.8\}$. For the assumed antibunching value of $g^{(2)}(0)=0.02$, an optimization of $\mu$ would always yield the largest allowed value, because the low antibunching value already suppresses multi-photon contributions and does not need to be balanced with the single-photon emission probability. In this case, increasing the mean photon number systematically improves the performance of the QD implementation, resulting in both higher secret key rates and a larger maximum transmission loss. Since a larger mean photon number also yields a higher probability of single-photon emission while maintaining a low multi-photon probability due to a relatively low antibunching value, the source approaches the behaviour of an ideal deterministic single-photon source. As a reference, we also consider QD implementations without decoy states. This includes the case of a perfect QD single-photon source without multi-photon emissions, which provides a benchmark for the attainable performance with ideal single-photon sources. Additionally, we consider the QD implementation without decoy states using the GLLP method \cite{GLLP:2004}, which makes the pessimistic assumption that all multi-photon signals emitted by Alice and Bob are received by the measurement unit. Notably, QD sources using decoy states outperform the optimised WCP protocol across all distances already for $\mu = 0.3$, despite the latter being optimised at every transmission loss.

To determine the minimum mean photon number $\mu$ required for the QD implementation to outperform the optimized WCP implementation, we plot the difference between the two secret key rates as a function of the transmission loss and the mean photon number $\mu$ of the QD source in Fig.~\ref{fig:MDI_WCP_QD_2D}. In general, the threshold $\mu$ depends on several parameters, including the transmission loss, channel and detector errors, and the characteristics of the employed sources. For the specific parameters considered here, we observe that the threshold mean photon number is around $\mu_{\mathrm{QD}} \approx 0.26$ in the low-loss regime and decreases with more loss to $\mu_{\mathrm{QD}} \approx 0.2$. This behavior can be understood from the optimization of the WCP mean photon number: in the low-loss regime, the optimal mean photon number for the WCP implementation is higher, and hence the QD implementation only outperforms it for sufficiently large $\mu_{\mathrm{QD}}$. 

\begin{figure}[H]
    \centering
\includegraphics[width=0.6\columnwidth]{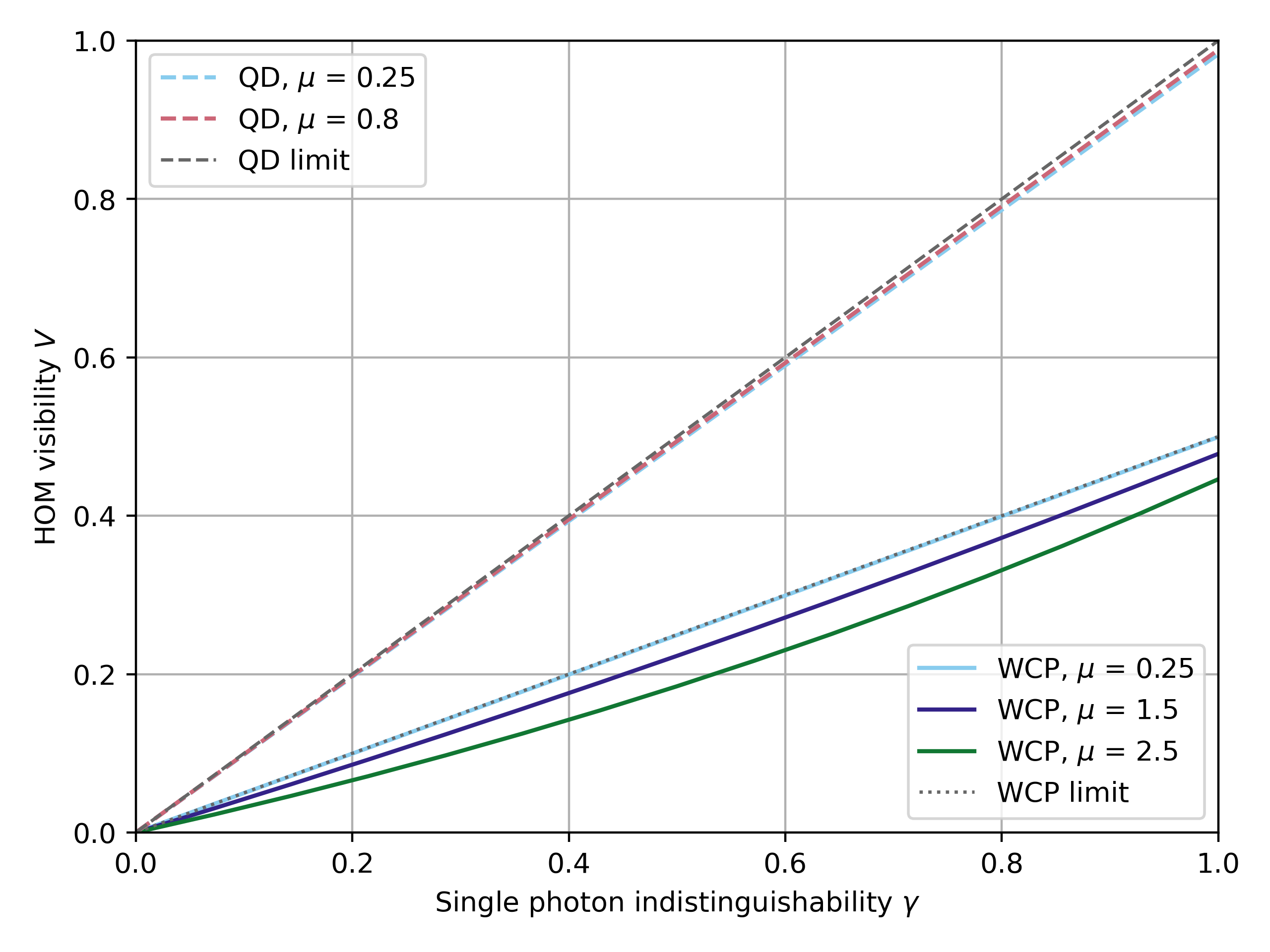}
    \caption{HOM visibility as a function of single-photon indistinguishability for different mean photon numbers, for WCP and QD sources. The black curves indicate the ideal limits for each source: the linear limit $V=\gamma/2$ for WCP and the ideal single-photon limit with perfect antibunching ($g^{(2)}(0)=0$) for QD sources. For the WCP, multi-photon contributions up to 15th order are included.}
    \label{fig:V_Vsp_WCP_QD}
\end{figure}

The results in Fig.~\ref{fig:MDI_WCP_QD} and ~\ref{fig:MDI_WCP_QD_2D} assume perfect photon indistinguishability ($\gamma = 1$), which represents an idealised scenario. In practice however, photon indistinguishability is always reduced to some degree, and its effect enters the security analysis through the HOM visibility $V$ - the experimentally accessible quantity that directly reflects the degree of two-photon interference at the MU. To connect the theoretical parameter $\gamma \in [0,1]$ to this observable, Fig.~\ref{fig:V_Vsp_WCP_QD} shows the HOM visibility as a function of the single-photon indistinguishability for both WCP and QD sources at different mean photon numbers (see Section~\ref{HOM_and_indistinguishability} for the underlying framework). For WCPs operated at low mean photon numbers, the visibility depends approximately linearly on $\gamma$, as previously shown~\cite{Duplinskiy:2021}. As the mean photon number increases, however, higher-order photon-number contributions become increasingly important and the dependence becomes non-linear. For QD sources, considerably higher HOM visibilities can be achieved due to their sub-Poissonian photon-number statistics. Nevertheless, increasing the mean photon number also increases the probability of multi-photon events, resulting in a reduction of the maximum achievable pulse visibility for a given value of $\gamma$.

\begin{figure}[htb]
    \centering
    \begin{subfigure}[t]{0.49\columnwidth}
        \centering
        \includegraphics[width=\linewidth]{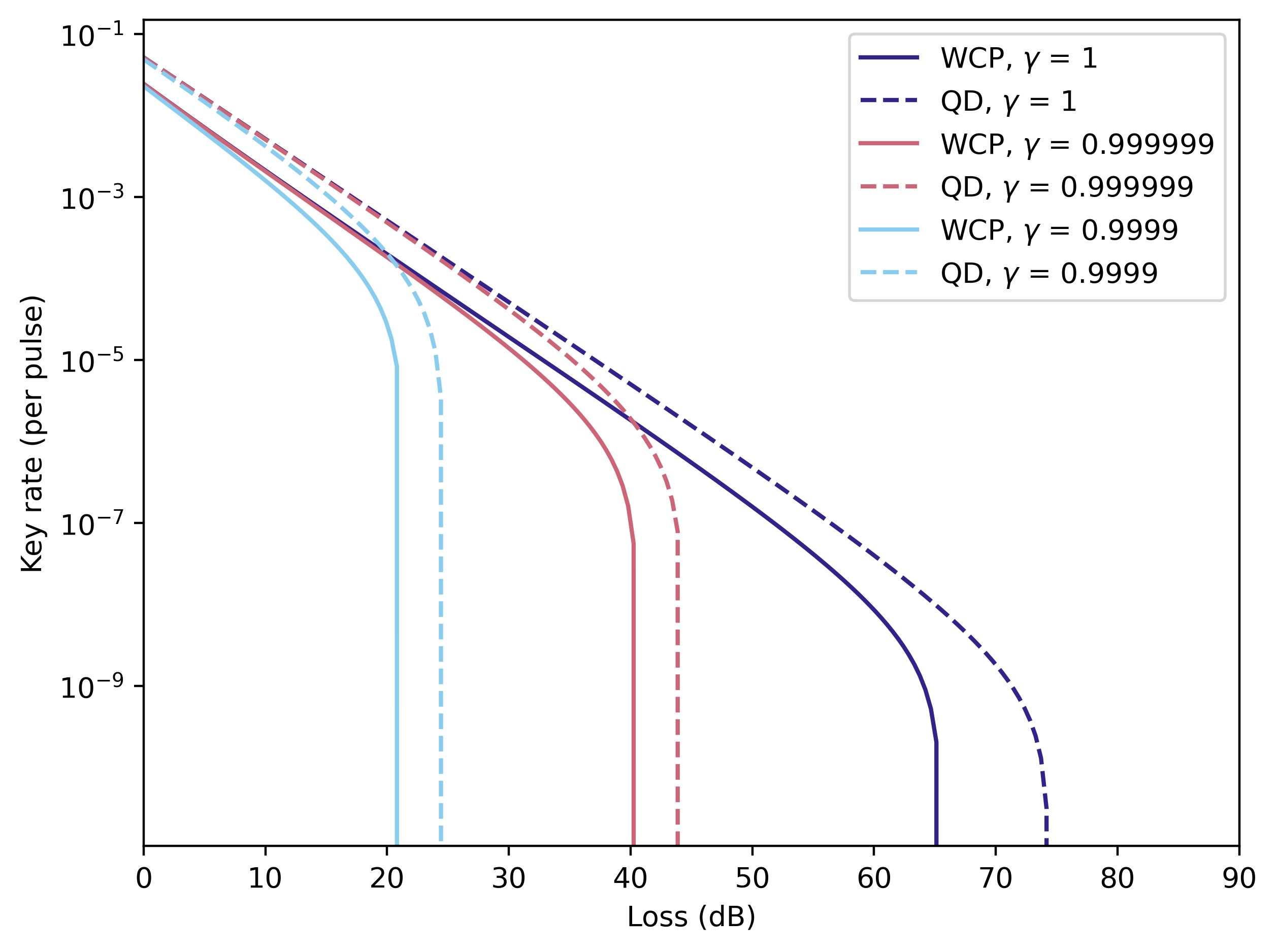}
        \caption{}
        \label{fig:MDI_WCP_QD_reduced}
    \end{subfigure}
    \hfill
    \begin{subfigure}[t]{0.49\columnwidth}
        \centering
        \includegraphics[width=\linewidth]{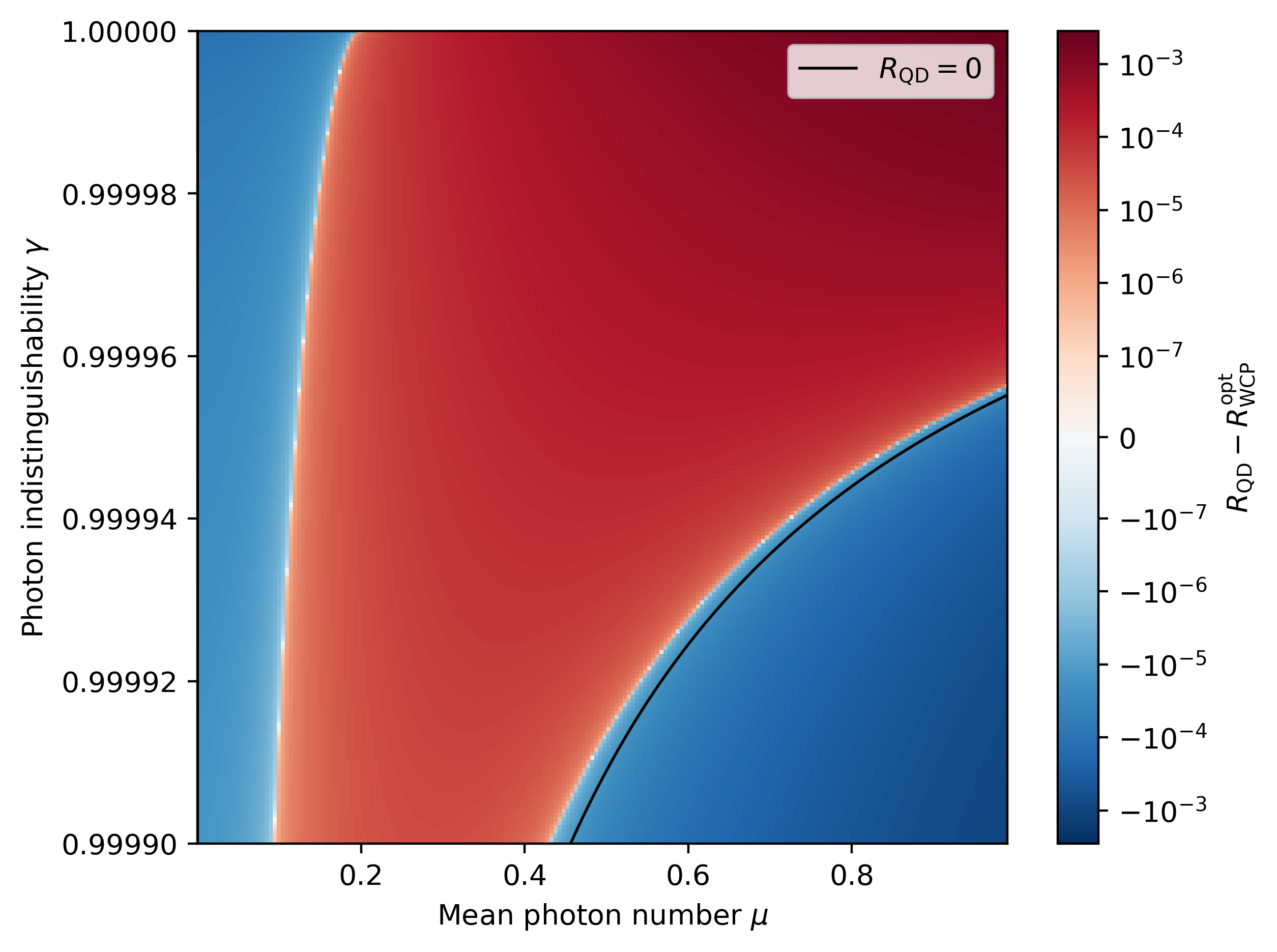}
        \caption{}
        \label{fig:MDI_mu_gamma_2D}
    \end{subfigure}
    \caption{
    Impact of imperfect single-photon indistinguishability on the asymptotic secret key rates of WCP- and QD-based MDI-QKD implementations. (a) Secret key rates as a function of the total loss between Alice and Bob for WCP (solid lines) and QD (dashed lines) sources with fixed mean photon number $\mu=0.3$, considering reduced single-photon indistinguishability. Fixing $\mu$ enables a direct mapping between the experimentally measurable HOM visibility $V$ and the predicted key rate. (b) Performance advantage of the QD source $R_{\mathrm{QD}}-R_{\mathrm{WCP}}^{\mathrm{opt}}$ as a function of the single-photon indistinguishability $\gamma$ and the mean photon number $\mu$ of the QD sources. The WCP mean photon number is numerically optimized, and the total loss is fixed at approximately $23\,\mathrm{dB}$, corresponding to a total distance between Alice and Bob of $30\,\mathrm{km}$ for the parameters in \cite{Lo:2012}.
    }
    \label{fig:MDI_WCP_QD_indistinguishability}
\end{figure}

Having established the relationship between $\gamma$ and the measurable HOM visibility $V$, we now incorporate imperfect indistinguishability directly into the security analysis. Figure~\ref{fig:MDI_WCP_QD_reduced} shows the resulting asymptotic key rates for both WCP and QD implementations at fixed mean photon number $\mu = 0.3$, for values of $\gamma$ close to the ideal case. For $\gamma=0.9999$, the HOM visibility for WCP is 0.499015, and for QD it is 0.983297, while for $\gamma=0.999999$, it is 0.499065 and 0.983394 respectively. In contrast to Fig.~\ref{fig:MDI_WCP_QD}, the mean photon number is not optimized for the WCP implementation, in order to relate the graph to a single HOM visibility. This enables a direct mapping between the experimentally measured HOM visibility and the predicted secret key rate. 

Even for single-photon indistinguishabilities very close to the ideal value ($\gamma=1$), e.g. $\gamma=0.999999$ the performance of both implementations is noticeably reduced compared to the ideal case. A further reduction to $\gamma=0.9999$ leads to a substantial decline of the achievable key rate and maximum transmission loss. In this case, secret key generation is possible only up to approximately $21\,\mathrm{dB}$ for WCP sources and $25\,\mathrm{dB}$ for QD sources. Using the fiber attenuation and detection efficiency of Ref.~\cite{Lo:2012}, this corresponds to transmission distances of approximately $25\,\mathrm{km}$ and $40\,\mathrm{km}$, respectively. These results highlight a critical sensitivity of MDI-QKD to even minute reductions in single-photon indistinguishability.

The strong sensitivity to photon indistinguishability raises the question of how this dependence affects the results across a broader parameter space. Therefore, we extend the analysis by considering both the photon indistinguishability $\gamma$ and the mean photon number $\mu$ of the QD source. Figure~\ref{fig:MDI_mu_gamma_2D} illustrates that the advantage of the QD implementation over the optimized WCP implementation depends jointly on the source brightness and photon indistinguishability. Increasing $\mu$ generally improves the QD performance, but only if the photon indistinguishability remains sufficiently high. Conversely, even a highly indistinguishable QD source does not necessarily outperform the optimized WCP implementation at sufficiently low $\mu$. Thus, an advantage over WCP sources requires a suitable combination of high source brightness and photon indistinguishability.


\section{Discussion}
\label{discussion}
In this work, we have investigated MDI-QKD implementations based on QD single-photon sources and benchmarked them against WCP systems. A key result is that QD-based implementations already outperform WCP ones at moderate mean photon numbers, despite the latter being evaluated under loss-dependent optimisation. This advantage stems directly from the sub-Poissonian photon-number statistics of QD sources: by strongly suppressing multi-photon contributions through a low $g^{(2)}(0)$, QD sources avoid the fundamental trade-off between maximising single-photon emission and suppressing multi-photon events that inherently limits WCP-based implementations. As a result, QD sources can be operated at larger mean photon numbers without incurring the multi-photon penalty that constrains WCP performance, leading to both higher key rates and extended maximum transmission distances.

Beyond photon statistics, we have incorporated the effect of imperfect single-photon indistinguishability into the security analysis, an aspect not previously considered in the MDI-QKD context. Our results reveal a striking sensitivity of both source types to even minute deviations from ideal indistinguishability. This behaviour reflects the fundamental role of HOM interference in MDI-QKD, and is further amplified compared to standard BB84~\cite{Duplinskiy:2021}, since indistinguishability is required between two independent sources rather than one.

To connect these findings to experimentally achievable performance, we note that state-of-the-art QD sources have recently demonstrated raw HOM visibilities of $V = 0.928$,  for a single QD emitter, corresponding to a corrected single-photon indistinguishability of $\gamma = 0.9856$~\cite{ding2025high}. Obtaining similarly high or higher HOM visibilities also for remote, spatially separated QD sources, as required in MDI- and twin-field QKD, is still a major challenge. To date, the highest value achieved for QD sources is $V_{\rm{remote}} = 0.930\pm0.008$~\cite{Zhai:2022}, which is not far from the values considered here, suggesting that future QD sources could well reach the predicted secret key rates for MDI-QKD. For WCP sources, HOM visibilities of $V = 0.499 \pm 0.004$ have been experimentally demonstrated~\cite{Comandar:16}, showcasing that WCP sources are already operating near their fundamental limit. However, we note here that typical values for HOM visibilities are usually much lower \cite{Pont:2024}. Therefore, our comparison underscores that further improvements in single-photon indistinguishability will play a critical role in unlocking the full performance advantage of QD-based MDI-QKD. That said, the measured HOM visibility also depends on imperfections in the measurement apparatus and therefore does not necessarily reflect the intrinsic indistinguishability of the emitted photons. If these imperfections are well characterised in a trusted and well calibrated measurement setup, it may be possible to infer the intrinsic indistinguishability more accurately and use this estimate in the security analysis. This could potentially mitigate the substantial key-rate penalties predicted by our analysis, provided that the inference can be rigorously justified within the security model.

Several aspects of our analysis merit further investigation. First, we have restricted our treatment to the asymptotic regime, in which an infinite number of signals is assumed. In practice, finite-size effects arising from statistical parameter estimation with a finite number of rounds and decoy states will reduce the achievable key rate and maximum transmission distance. These effects are expected to impact WCP and QD implementations differently, since the two source types have distinct photon-number distributions and therefore different statistical fluctuations in the decoy-state analysis \cite{Morrison:2023}. Second, we have assumed symmetric channel conditions, with Alice and Bob placed equidistant from the MU, as well as identical source characteristics for both users (i.e. identical $g^{(2)}(0)$ and $V$). Asymmetric configurations, relevant for deployed network scenarios, may further differentiate the performance of the two source types and warrant dedicated study. Third, while we have fixed $g^{(2)}(0) = 0.02$ throughout, consistent with recent experimental demonstrations~\cite{ding2025high}, further reductions in multi-photon emission, down to $g^{(2)}(0) \sim 10^{-4}$ ~\cite{Schweickert:2018,hanschke2018quantum}, would push QD performance even closer to the ideal deterministic single-photon source limit, provided that high indistinguishability can be maintained simultaneously. Addressing these aspects in future work will be essential for a complete assessment of QD single-photon sources as a practical platform for long-distance MDI-QKD.

\section{Methods}
\label{methods}
\subsection{Asymptotic key rate}
\label{key_rate}
In order to compute the secret key rate for MDI-QKD that is achieved using the decoy-state method for the two photon sources under consideration, we follow previous studies \cite{Lo:2012,Lo:2005_decoy}. We first model the transmission of photons from one of the communicating parties to the measurement unit over a fiber of length $\ell/2$ (measured in km). The probability that a single photon is successfully transmitted is given by
\begin{align*}
\eta(\ell) = P_{\mathrm{d}} \cdot 10^{-\frac{\beta\, \ell}{2\cdot 10}},
\label{eta}
\end{align*}
where $P_{\mathrm{d}}$ accounts for all distance-independent losses, including fiber-to-relay coupling and detector inefficiencies, and $\beta$ denotes the fiber attenuation coefficient in units of dB per km. Throughout this work, use $P_{\mathrm{d}}=0.145$ and $\beta=0.2,\mathrm{dB/km}$ \cite{Lo:2012}. Alternatively, the dependence can be expressed in terms of the total loss between Alice and Bob in dB, including all distance-dependent and distance-independent losses. Assuming that the total loss is equally distributed between the two links, the single-link transmission probability is given by $\eta = 10^{-\frac{\mathrm{dB}}{2\cdot 10}}$. For an $n$-photon state, the probability that at least one photon is transmitted is
$
\eta_n = 1 - (1 - \eta)^{n}.
$
The \emph{yield} $Y_{nm}$, defined as the probability of a successful detection of an $n$-photon state from Alice and an $m$-photon state from Bob, can then be expressed as
\begin{align*}
Y_{nm} =
\left( P_{\mathrm{dc}} + \eta_n - P_{\mathrm{dc}}\, \eta_n \right) \left( P_{\mathrm{dc}} + \eta_m - P_{\mathrm{dc}}\, \eta_m \right).
\end{align*}
where $P_{\mathrm{dc}}$ is the probability of a detector dark count, which we set to $P_{\mathrm{dc}}=6.02\cdot 10^{-6}$ \cite{Lo:2012}. 

Depending on the type of light source employed by Alice and Bob, we obtain a different photon-number distribution $P_{nm}$. WCPs follow the Poisson distribution; therefore, for two independent sources operated by Alice and Bob, the probability that Alice's pulse contains $n$ photons and Bob's pulse contains $m$ photons is given by
\begin{align*}
P^{\mathrm{WCP}}_{nm}
= e^{-\mu_A} \frac{\mu_A^{n}}{n!}
  \; e^{-\mu_B} \frac{\mu_B^{m}}{m!}.
\end{align*}
where $\mu_A$ and $\mu_B$ are the mean photon numbers of the two sources, respectively.

For QD sources, emission statistics differ significantly \cite{Waks:2002}. The probabilities that a single QD pulse contains $i\in\{0,1,2\}$ photons are given by
\begin{align*}
P^{\mathrm{QD}}_{1} &= \mu~,~~~~~~~
P^{\mathrm{QD}}_{2} = \frac{1}{2} \mu^2 g^{(2)}(0)~,~~~~~~~ 
P^{\mathrm{QD}}_{0} = 1 - P^{\mathrm{QD}}_{1} - P^{\mathrm{QD}}_{2},
\end{align*}
where $\mu$ denotes the single-photon emission probability of the QD source and $g^{(2)}(0)$ the antibunching parameter of the quantum dot. Unless stated otherwise, we set $g^{(2)}(0)=0.02$. The probability that Alice's pulse contains $n$ photons and Bob's pulse contains $m$ photons with $n, m \in \{0, 1, 2\}$ is again $P^{\mathrm{QD}}_{nm} = P^{\mathrm{QD}}_n P^{\mathrm{QD}}_m$.

We can now characterize the output-to-input ratio for $n$ and $m$ photons in Alice's and Bob's pulses, respectively, by defining the gain $Q_{nm}=P_{nm}\, Y_{nm}$. The \emph{overall gain} can be expressed as
\begin{align*}
Q = \sum_{n=0}^{\infty} \sum_{m=0}^{\infty} Q_{nm} = \sum_{n=0}^{\infty} \sum_{m=0}^{\infty} P_{nm}\, Y_{nm}.
\end{align*}

\noindent The asymptotic secret key rate for MDI-QKD using the decoy-state method \cite{Lo:2012,Lo:2005_decoy} is given by
\begin{align*}
R = Q^Z_{11}\left[1 - h(e^X_{11})\right] - Q^Z\, f\, h(e^Z),
\end{align*}
where the superscripts $X$ and $Z$ indicate the specific type of measurement performed, $h(\cdot)$ the binary entropy function and $f$ the error-correction inefficiency, which we set to $f=1.16$ \cite{Lo:2012}. $e$ denotes the overall quantum bit error rate (QBER), weighted by the gain of each $(n,m)$ photon-number pair:
\begin{align*}
e
= \frac{1}{Q}
\sum_{n=0}^{\infty}\sum_{m=0}^{\infty}
Q_{nm}\, e_{nm}.
\end{align*}
where each individual $e_{nm}$ incorporates contributions from measurement device misalignment - characterized by the misalignment probability $e_{\mathrm{d}}$, which is assumed to be $e_{\mathrm{d}}=0.015$ \cite{Lo:2012} - as well as dark counts: 
\begin{align*}
e_{nm} =
\frac{1}{Y_{nm}}
\Biggl(
2 e_{\mathrm{d}}\,\eta_n\eta_m
+ \frac{1}{2} P_{\mathrm{dc}} \left( \eta_m + \eta_n \right)
+ \frac{1}{4} P_{\mathrm{dc}}^{2}
\Biggr).
\end{align*}

\subsection{Hong-Ou-Mandel interference and single-photon distinguishability}
\label{HOM_and_indistinguishability}
Photon indistinguishability is a key requirement for the security of MDI-QKD, as the Bell-state measurement at the measurement unit succeeds only if the two incoming photons are indistinguishable. If that is not the case, the fidelity of the measurement degrades, opening side-channel vulnerabilities that an attacker could exploit. A standard method to quantify indistinguishability is the HOM interference experiment, in which two optical pulses interfere on a balanced beam splitter and the coincidence events at the two output ports are recorded. A relative temporal delay $\Delta t$ between the input pulses is introduced to control their distinguishability. For sufficiently large $\Delta t$, the photons do not overlap temporally and behave independently. Hence, half of the two-photon events yield coincidence clicks. In contrast, when $\Delta t$ is reduced such that the photons overlap, the coincidence rate decreases due to two-photon interference. In the ideal case of perfectly indistinguishable single photons, the coincidence events vanish completely.

The degree of HOM interference is commonly quantified by the visibility, defined as
\begin{align}
    V = \frac{N_{\mathrm{max}} - N_{\mathrm{min}}}{N_{\mathrm{max}}},
\end{align}
where $N_{\mathrm{max}}$ and $N_{\mathrm{min}}$ denote the coincidence counts for fully distinguishable (large $\Delta t$) and maximally indistinguishable photons ($\Delta t = 0$), respectively. There are fundamental limits to the achievable HOM visibility for different photonic sources. To illustrate this, consider two perfectly indistinguishable pulses with identical photon-number statistics
\begin{align*}
\rho = \sum_{n=0}^{\infty} P_n \ket{n}\!\bra{n},
\end{align*}
where $P_n$ denotes the probability of emitting $n$ photons. The joint input state at the beam splitter for two pulses is then given by
\begin{align}
\rho' =\,& \rho_1 \otimes \rho_2 \notag \\
=\,& P_0^2 \ket{0_1 0_2}\!\bra{0_1 0_2}
+ P_0P_1 \ket{1_1 0_2}\!\bra{1_1 0_2}
+ P_0P_1 \ket{0_1 1_2}\!\bra{0_1 1_2}
+ P_1^2 \ket{1_1 1_2}\!\bra{1_1 1_2} \notag \\
&+ P_0P_2 \ket{2_1 0_2}\!\bra{2_1 0_2}
+ P_0P_2 \ket{0_1 2_2}\!\bra{0_1 2_2}
+ \mathcal{O}\left(N_p\!>\!2\right),
\label{input_BS}
\end{align}
where $\mathcal{O}(N_p>2)$ summarizes all contributions with total photon number $N_p>2$. Vacuum and single-photon contributions do not lead to coincidence events and therefore do not contribute to the HOM visibility. The term $\ket{1_1 1_2}\!\bra{1_1 1_2}$ yields perfect two-photon interference $V=1$. In contrast, the two-photon states $\ket{2_1 0_2}\!\bra{2_1 0_2}$ and $\ket{0_1 2_2}\!\bra{0_1 2_2}$ have both photons entering from the same input port. In this case, the beam splitter output is a superposition in which both photons can exit via either port with equal probability, yielding a coincidence rate of $50\%$ regardless of indistinguishability. Therefore, these terms do not contribute to the visibility ($V=0$).

For WCP, the Poissonian photon-number distribution satisfies $P_1^2 = 2P_0P_2$. Thus, when neglecting higher order terms $N_p > 2$, one obtains the well-known upper bound $V_{\mathrm{WCP}} = 0.5$ even for perfectly indistinguishable photons~\cite{Mandel:1983}. However, this result relies on the low-intensity approximation ($\mu \ll 1$). For larger mean photon numbers, or for QD sources with non-negligible multi-photon emission characterized by a finite $g^{(2)}(0)$, higher-order Fock-state contributions must be taken into account. We describe each input pulse by a density operator in the Fock basis as $\rho_1$ and $\rho_2$ with
\begin{align*}
\rho_1 = \sum_{n=0}^{\infty} P_n \ket{n}_1\!\bra{n}_1 = \sum_{n=0}^{\infty} \frac{P_n}{n!} (a^{\dagger}_1)^n \ket{0}\!\bra{0} (a_1)^n,
\end{align*}
where $P_n$ denotes the probability of emitting $n$ photons. We introduce partial distinguishability by decomposing the creation operator of the first  input mode into components parallel and orthogonal to the second mode,
\begin{align*}
a_2^\dagger = \sqrt{\gamma}\,a_1^\dagger + \sqrt{1-\gamma}\,a^\dagger_{\perp1},
\end{align*}
where $a_\perp^\dagger$ creates photons in a mode orthogonal to $a^\dagger$ and $\ket{n}_2 = \frac{1}{\sqrt{n!}}\left(a_2^\dagger\right)^n \ket{0}$ denotes the $n$-photon Fock state in the second mode. The parameter $\gamma\in[0,1]$ quantifies the single-photon indistinguishability, with $\gamma=1$ corresponding to perfectly indistinguishable photons and $\gamma=0$ to fully distinguishable photons. 

The HOM visibility is obtained by evaluating the joint state $\rho'=\rho_1\otimes\rho_2$, and summing up the coincidence probabilities over all photon-number contributions. Ref.~\cite{Duplinskiy:2021} further derived an analytical expression for the fidelity between the states $\rho_1$ and $\rho_2$ as a function of the single-photon indistinguishability $\gamma$:
\begin{align*}
\sqrt{F(\rho_1,\rho_2)} = P_0 + \sum_{n=1}^{\infty} P_n \gamma^{n/2},
\end{align*}
where the fidelity between two states is defined as $F(\rho_1, \rho_2)=\left(\text{Tr} \lvert \sqrt{\rho_1}\sqrt{\rho_2}\rvert\right)^2$. This expression can be rewritten for the different photon sources with the respective photon number statistics:
\begin{align}
\text{WCP}&:&\sqrt{F(\rho^{\mathrm{WCP}}_1,\rho^{\mathrm{WCP}}_2)} &= \exp\!\left[\mu\left(\sqrt{\gamma}-1\right)\right] \label{fid_gamma_WCP}\\
\text{QD}&:&\sqrt{F(\rho^{\mathrm{QD}}_1,\rho^{\mathrm{QD}}_2)} &= P_0^{\mathrm{QD}} + P_1^{\mathrm{QD}}\sqrt{\gamma} + P_2^{\mathrm{QD}}\gamma.\label{fid_gamma_QD}
\end{align}

\subsection{Impact of imperfect photon indistinguishability on key rate}
\label{HOM_correction}
Ideally, an external eavesdropper should be unable to distinguish between the quantum states prepared by Alice and Bob in the $X$ and $Z$ bases, since the corresponding density matrices
\begin{align*}
\rho_Z = \frac{1}{2}\ket{0}\!\bra{0} + \frac{1}{2}\ket{1}\!\bra{1},~~~
\rho_X = \frac{1}{2}\ket{+}\!\bra{+} + \frac{1}{2}\ket{-}\!\bra{-},
\end{align*}
are identical for perfectly prepared states. In practice, however, experimental imperfections, such as reduced photon indistinguishability, break this symmetry, rendering the two bases partially distinguishable and potentially allowing an eavesdropper to gain information. We follow the framework of~\cite{Duplinskiy:2021} to incorporate such imperfections into the security analysis of MDI-QKD, quantifying their effect through the \emph{basis imbalance}. For the BB84 protocol, this parameter is given by~\cite{Koashi:2009, Lucamarini:2015}
\begin{align}
    \Delta = \frac{1 - \sqrt{F(\rho_X^A, \rho_Z^A)}}{2},
\end{align}
where $F(\rho_X^A, \rho_Z^A)$ denotes the fidelity between the density matrices in the $X$ and $Z$ bases emitted by Alice. 
For MDI-QKD, this expression must be extended to account for the fact that both Alice and Bob prepare quantum states~\cite{Tamaki:2012}. In this case, the basis imbalance reads
\begin{align}
    \Delta = \frac{1 - \sqrt{F(\rho_X^A, \rho_Z^A)} \sqrt{F(\rho_X^B, \rho_Z^B)}}{2}.
    \label{basis_imbalance_MDI}
\end{align}
This expression includes the product of two fidelities (for Alice's and Bob's pulses) and therefore scales worse than the BB84 expression above.

Ref.~\cite{Duplinskiy:2021} derives a lower bound on the fidelity between the $X$- and $Z$-basis density matrices in terms of the fidelities associated with the non-operational degrees of freedom. The derivation is based on the decomposition of the quantum states into their operational and non-operational components and the application of the triangle inequality for the Bures angle~\cite{Ma:2009}. We refer the reader to Ref.~\cite{Duplinskiy:2021} for the full derivation. In general, the bound depends on the fidelities of all pairs of prepared states and therefore allows for different distinguishabilities between individual pulse pairs. Throughout this work, however, we assume that all pulse pairs are characterized by the same fidelity $F$. Under this assumption, the bound simplifies to
\begin{align}
\sqrt{F(\rho_X, \rho_Z)}
\geq
\cos\Biggl[
\cos^{-1}\!\left(\sqrt{F}\right)
+ 2\cos^{-1}\!\left(\frac{1+\sqrt{F}}{2}\right)
\Biggr].
\label{eq:Fxz_bound}
\end{align}
This expression can be used to quantify the basis imbalance introduced in Eq.~\eqref{basis_imbalance_MDI}.
For WCP and QD sources, $\sqrt{F}$ is given by Eqs.~\eqref{fid_gamma_WCP} and ~\eqref{fid_gamma_QD} respectively. Effectively, the basis imbalance increases the single-photon error rate \cite{Tamaki:2012, Lucamarini:2015} and in that way reduces the overall performance of the protocol. The modified error rate is given by\cite{Lo:2007, Koashi:2009}
\begin{align}
    e'_{11} = e_{11} + 4 \Delta' (1 - \Delta') (1 - 2 e_{11})+ 4 (1 - 2 \Delta') \sqrt{\Delta' (1 - \Delta') e_{11} (1 - e_{11})},
    \label{e11_prime}
\end{align}
where $e_{11}$ denotes the single-photon error rate, and $\Delta'$ is defined as
\begin{align}
    \Delta' = \min \left\{ \frac{\Delta}{Y_{11}}, \frac{1}{2} \right\},
    \label{delta_prime}
\end{align}
with $Y_{11}$ being the single-photon yield. Eq.~\eqref{delta_prime} also implies that the correction becomes increasingly significant at larger transmission distances, since the single-photon yield $Y_{11}$ decreases exponentially with the distance between the measurement unit and Alice and Bob. $\Delta'$ is capped at $\frac{1}{2}$ to ensure that the last correction term in Eq.~\eqref{e11_prime} remains non-negative and thus does not reduce the single-photon error rate.
\section{Data Availability}
All data are available from the corresponding author upon reasonable
request.
\section{Acknowledgements}
J.M. and A.P. acknowledge funding via the Quant-GPlCz Project (supported by BMDV grant No. 19OS25001A and the EU’s
Horizon Europe Programme grant No. 101249338), QIA Phase 1 (supported by the EU's Horizon Europe research and innovation Programme grant No. 101102140) and Berlin Quantum. N. W. acknowledges support from the BMFTR (QR.N). T.H. further acknowledges financial support from the BMFTR through the joint projects “tubLAN Q.0” (Grant No. 16KISQ087K) and “QuNET+ICLink” (Grant No. 16KIS1967), as part of the federal government’s IT security research framework “Digital. Secure. Sovereign.”. \\
\section{Author contributions}
A.P. and T.H. conceived the project. A.P. supervised the project. J.M. carried out the calculations and simulations, with contributions from Y.L. and N.W. J.M. wrote the initial manuscript draft, and all authors contributed to manuscript review and editing.
\section{Competing interests}
The authors declare no competing interests.
\newpage
 
\bibliographystyle{ieeetr}
\bibliography{MDI_bib}

\appendix

\end{document}